\documentclass[12pt,a4paper]{article}

\usepackage{jheppub}
\makeatletter
\gdef\@fpheader{} %
\makeatother
\usepackage{amsthm}
\usepackage{slashed}
\usepackage{array}
\usepackage[perpage,symbol,flushmargin]{footmisc}
\everymath{\displaystyle}
\renewcommand{\thetable}{\Roman{table}}
\makeatletter %
\renewcommand{\affiliation}[2][]{%
\affiltrue
  \if!#1!%
    \affil@toks=\expandafter{\the\affil@toks{\item[]\footnotesize #2}}%
  \else
    \affil@toks=\expandafter{\the\affil@toks{\item[$^{#1}$]\footnotesize #2}}%
  \fi
}
\renewcommand{\@email}[1]{\href{mailto:#1}{\ttfamily\footnotesize #1}}%
\makeatother

\title{Extremal non-rotating black holes have no fermionic Love}

\author[a,*]{Xiankai Pang,\note[*]{Corresponding author.}}
\author[b,c]{Yu Tian,}
\author[d,e]{and Hongbao Zhang}
\affiliation[a]{School of Physics and Astronomy, China West Normal University, Nanchong 637009, China}
\affiliation[b]{School of Physical Sciences, University of Chinese Academy of Sciences, Beijing 100049, China}
\affiliation[c]{Institute of Theoretical Physics, Chinese Academy of Sciences, Beijing 100190, China}
\affiliation[d]{School of Physics and Astronomy, Beijing Normal University, Beijing 100875, China}
\affiliation[e]{Key Laboratory of Multiscale Spin Physics, Ministry of Education, Beijing Normal University, Beijing 100875, China}
\emailAdd{xkpang@cwnu.edu.cn}
\emailAdd{ytian@ucas.ac.cn}
\emailAdd{hongbaozhang@bnu.edu.cn}

\abstract{The static tidal Love numbers (TLNs) of $4$-dimensional black holes
vanish for bosonic perturbations but are generically nonzero for fermions, with
rare exceptions. In this paper, we show that for static, spherically symmetric
black holes, fermionic TLNs vanish if and only if the black hole is extremal, in
the sense that its horizon is degenerate. This follows from a closed formula for
the static fermionic TLN of any asymptotically flat black hole, obtained by
solving the static massless Dirac equation exactly on an arbitrary such spacetime
and imposing regularity at the horizon. As applications, we analyze the
Culetu--Simpson--Visser regular black hole and the loop-quantum-gravity remnant black
holes, whose extremal configurations lead to vanishing fermionic yet nonvanishing
bosonic TLNs.}

\keywords{Fermionic Love number; Extremal black hole; Spherically symmetric spacetime; Tidal deformation}

\begin{document}

\maketitle

\section{Introduction}

When a self-gravitating body is placed in an external tidal field, it develops
induced multipole moments; their ratios to the applied field define the body's
tidal Love numbers (TLNs) \cite{Love1909}. In relativistic gravity, TLNs quantify the conservative
tidal response of compact objects
\cite{Goldberger:2004jt,Damour:2009va,Binnington:2009bb,Damour:2009vw,Fang:2005qq,Bini:2012gu,Porto:2016pyg,Chakraborty:2026qru}
and imprint directly on the gravitational-wave signal of compact binaries
\cite{Hinderer:2007mb,Flanagan:2007ix,Damour:2012yf,Yagi:2013bca,Yagi:2016ejg,LIGOScientific:2018cki}.
A striking property of four-dimensional black holes in general relativity is that
their static TLNs vanish identically for scalar, electromagnetic, and gravitational
perturbations
\cite{Binnington:2009bb,Damour:2009vw,Gurlebeck:2015xpa,LeTiec:2020bos,Chia:2020yla,Poisson:2020mdi}.
Over the past few years, this vanishing has been understood as the consequence of a
hidden \emph{ladder symmetry} of the static perturbation equations
\cite{Hui:2020xxx,Hui:2021vcv,Berens:2022ebl,Charalambous:2021mea,Ivanov:2022qqt},
whose geometric origin is an approximate anti-de Sitter$_2$ (AdS$_2$) isometry of the $(t,r)$ geometry
\cite{Charalambous:2021mea,Berens:2025okm}. Demanding the ladder symmetry
constrains the background. The most general static, spherically symmetric black
hole spacetime admitting it for scalar perturbations is spatially conformal to
Reissner--Nordstr\"om (RN), i.e.\ of the form
\begin{equation}\label{eq:scalargfamily}
ds^2 = \frac{\Delta_b}{h}\,dt^2 - \frac{h}{\Delta_b}\,dr^2 - h\,d\Omega_2^2\,,\qquad
\Delta_b = r^2 - c_2 r + c_3
\end{equation}
with $c_2,c_3$ constants, and the symmetry is necessary as well as sufficient for
the vanishing \cite{Sharma:2024hlz,Sharma:2025xii}.

Nonzero TLNs, however, are just as widespread. They have by now been found for
time-dependent (dynamical) tides
\cite{Chakraborty:2023zed,Perry:2023wmm,Chakraborty:2025wvs,HegadeKR:2025qwj,Chakraborty:2026dox,Apostolidis:2026qsg},
in higher dimensions \cite{Kol:2011vg,Hui:2020xxx,Cardoso:2019vof,Pang:2026vah}, in
AdS backgrounds \cite{Franzin:2024cah}, for magnetically charged black
holes \cite{Pereniguez:2025jxq}, in modified gravity and for exotic compact objects
\cite{Cardoso:2017cfl,Cardoso:2019rvt,Maggio:2021ans,DeLuca:2022tkm}, and in quantum-corrected
spacetimes \cite{Barbosa:2025uau,Motaharfar:2025typ,Barbosa:2026qcv}. Most recently, the static response to massless
fermionic fields was shown to be generically nonzero as well. Computations on
Schwarzschild and Kerr \cite{Chakraborty:2025zyb} and on RN
\cite{Pang:2025myy} black holes found \emph{nonzero} fermionic TLNs, which for
RN vanish only in the extremal limit.

This pattern raises two natural questions. First, on precisely which geometries do
fermionic TLNs vanish? Can the answer be characterized as sharply as in the
bosonic case, where the vanishing is tied to the conformal-RN family
\eqref{eq:scalargfamily}? Second, when the fermionic response does vanish, what
happens to the scalar response on the same background?

In this paper we address these two questions. As we will see, however, and unlike
the bosonic case, the fermionic ladder does not constrain the metric
functions at all (appendix~\ref{app:ladder}); on the other hand, the static Dirac equation is exactly solvable
on arbitrary static, spherically symmetric spacetimes, which enables us to tie the
vanishing of fermionic TLNs to extremal black holes. Applied to the
Culetu--Simpson--Visser (CSV) regular black hole \cite{Culetu:2013fsa,Culetu:2014lca,Simpson:2019mud} and the
loop-quantum-gravity (LQG) remnant black holes of Borges et al.\
\cite{Borges:2023fub}, the fermionic TLNs indeed vanish at the
extremal configurations, while the scalar TLNs cannot vanish there, since
both families lie outside the conformal-RN class \eqref{eq:scalargfamily}. These
provide explicit examples of vanishing fermionic but nonvanishing bosonic TLNs.

A comment on observability is in order. Bosonic TLNs imprint on gravitational
waveforms and are constrained by observations
\cite{Flanagan:2007ix,Damour:2012yf,LIGOScientific:2018cki}. Fermionic TLNs have no
such channel. There is no macroscopic coherent fermionic tidal field, and a
fermionic environment of a black hole is intrinsically quantum. Our motivation is
structural. Fields of different spin probe the same background geometry through
wave operators of different order
\cite{Chandrasekhar:1985kt,Kokkotas:1999bd,Berti:2009kk}, and the static tidal
problem is the cleanest arena in which this difference shows up
\cite{Porto:2016zng}. In the present case, the static, first-order Dirac operator
reads only the conformal geometry of the spatial slice.

The paper is organized as follows. In section~\ref{sec:dirac} we derive the
Dirac equation for a general spherically symmetric spacetime in
Eddington--Finkelstein (EF) coordinates. The exact solution, the
horizon-regularity condition, and the resulting general TLN formula with its
checks are presented in section~\ref{sec:tln}. The vanishing criterion and its
physical realizations are the subject of section~\ref{sec:vanishing}.
Section~\ref{sec:conclusions} concludes.

Throughout the paper, we work in natural units $c=G=\hbar=1$ and in signature
$(+---)$, except where stated explicitly.

\section{The radial Dirac equation}
\label{sec:dirac}

\subsection{The general metric and the EF tetrad}

Consider the most general static, spherically symmetric spacetime, written in the
usual form
\begin{equation}\label{eq:metric}
ds^2 = f(r)\,dt^2 - \frac{dr^2}{g(r)} - h(r)\,d\Omega_2^2\,,
\end{equation}
with $d\Omega_2^2=d\theta^2+\sin^2\theta\,d\varphi^2$. We assume throughout that
the spacetime is asymptotically flat with power-law falloff,
\begin{equation}\label{eq:falloff}
f=1+O(r^{-p})\,,\qquad
g=1+O(r^{-p})\,,\qquad
h=r^2\big[1+O(r^{-p})\big]\,,\qquad p>0\,,
\end{equation}
implying in particular $gh/r^2=1+O(r^{-p})$. Since we will need to impose
regularity on the horizon, where the metric \eqref{eq:metric} is singular, we pass to the
ingoing EF coordinate
\begin{equation}\label{eq:efcoord}
v = t + r_*\,,\qquad \frac{dr_*}{dr}=\frac{1}{\sqrt{f(r)g(r)}}\,,
\end{equation}
in terms of which
\begin{equation}\label{eq:metricEF}
ds^2 = f(r)\,dv^2 - 2\sqrt{\frac{f(r)}{g(r)}}\,dv\,dr - h(r)\,d\Omega_2^2\,,
\end{equation}
which is regular at any non-degenerate horizon, where $f$ and $g$ share a simple
zero so that $\sqrt{f/g}$ is finite and nonzero\footnote{The integral
\eqref{eq:efcoord} converges at a non-degenerate horizon. At a degenerate horizon
it diverges, and the naive $v$ is not the right ingoing coordinate. Regular EF-type
coordinates exist there (the standard extremal-RN construction) but must be built
separately. Our regularity analysis is needed only at non-degenerate horizons. The
degenerate case is treated in section~\ref{sec:vanishing}, and the extremal TLN is taken as a limit of the non-degenerate
family.}.

A massless Dirac field decouples into two Weyl fields, $\nabla_{AA'}\psi^A=0$ and
its conjugate; we consider only $\psi^A$ here, and its conjugate gives the same
system with the roles of the two angular parities interchanged. We use the Newman--Penrose (NP) formalism \cite{Newman:1961qr}, and choose
the null tetrad
\begin{equation}\label{eq:tetrad}
l^\mu=\Big(1,\frac{\sqrt{fg}}{2},0,0\Big)\,,\qquad
n^\mu=\Big(0,-\sqrt{\frac gf},0,0\Big)\,,\qquad
m^\mu=\frac{1}{\sqrt{2h}}\Big(0,0,1,\frac{i}{\sin\theta}\Big)\,,
\end{equation}
with $l\cdot n=1$, $m\cdot\bar m=-1$, and all other inner products zero. This
generalizes the RN tetrad of \cite{Pang:2025myy} to $f\neq g$. Both $l^\mu$ and
$n^\mu$ are manifestly regular at any horizon where $f/g$ is
finite. The only nonvanishing spin coefficients are\footnote{Our spin-coefficient
conventions follow \cite{Pang:2025myy}.}
\begin{equation}\label{eq:spincoeff}
\rho=\frac{\sqrt{fg}\,h'}{4h}\,,\qquad
\varepsilon=-\frac14\sqrt{\frac gf}\,f'\,,\qquad
\mu=\sqrt{\frac gf}\,\frac{h'}{2h}\,,\qquad
\alpha=-\beta=\frac{\cot\theta}{2\sqrt{2h}}\,,
\end{equation}
where the prime $'$ denotes derivative respect to $r$.

\subsection{Weyl equation and separation of variables}

With the directional derivatives $D=l^\mu\partial_\mu$,
$\Delta=n^\mu\partial_\mu$, $\delta=m^\mu\partial_\mu$,
$\bar\delta=\bar m^\mu\partial_\mu$, the Weyl equation
$\nabla_{AA'}\psi^A=0$ for the dyad components $\psi^A=\psi^1 o^A+\psi^2\iota^A$
reads \cite{Pang:2025myy}
\begin{equation}\label{eq:weyl}
(D+\rho-\varepsilon)\psi^1+(\bar\delta+\alpha-\pi)\psi^2=0\,,\qquad
(\delta+\tau-\beta)\psi^1+(\Delta+\gamma-\mu)\psi^2=0\,.
\end{equation}
Substituting the tetrad \eqref{eq:tetrad} and the spin coefficients
\eqref{eq:spincoeff} gives the Weyl equation on the general metric
\eqref{eq:metricEF}
\begin{align}
\Big[\partial_v+\frac{\sqrt{fg}}{2}\partial_r
+\frac{\sqrt{fg}\,h'}{4h}+\frac14\sqrt{\frac gf}\,f'\Big]\psi^1
-\frac{1}{\sqrt{2h}}\,\bar\eth\,\psi^2&=0\,,\label{eq:weylA}\\
-\sqrt{\frac gf}\Big(\partial_r+\frac{h'}{2h}\Big)\psi^2
-\frac{1}{\sqrt{2h}}\,\eth\,\psi^1&=0\,,\label{eq:weylB}
\end{align}
where $\eth$ and $\bar\eth$ are operators on the two-sphere, acting on a
spin-weight-$s$ field $\eta_s$ as
$\eth\,\eta_s=-(\partial_\theta+i\csc\theta\,\partial_\varphi-s\cot\theta)\eta_s$ and
$\bar\eth\,\eta_s=-(\partial_\theta-i\csc\theta\,\partial_\varphi+s\cot\theta)\eta_s$
\cite{Pang:2025myy}.

We separate variables with the spin-weighted spherical harmonics
$S_\pm\equiv{}_{\pm1/2}Y_{lm}$ ($l=\frac12,\frac32,\dots$), which obey
\begin{equation}\label{eq:eth}
\eth\,S_-=\lambda\,S_+\,,\qquad
\bar\eth\,S_+=-\lambda\,S_-\,,\qquad
\lambda\equiv l+\frac12=1,2,3,\dots\,,
\end{equation}
and make the ansatz
\begin{equation}\label{eq:ansatz}
\psi^1=\mathrm{e}^{-i\omega v}\,\frac{R_-(r)}{\sqrt{h}\,\sqrt f}\,{}_{-1/2}Y_{lm}\,,\qquad
\psi^2=\mathrm{e}^{-i\omega v}\,\frac{R_+(r)}{\sqrt2\,\sqrt h}\,{}_{+1/2}Y_{lm}\,.
\end{equation}
The prefactors are chosen to absorb the spin-connection terms in
\eqref{eq:weylA}--\eqref{eq:weylB}. Substituting the ansatz
\eqref{eq:ansatz} into equations \eqref{eq:weylA}--\eqref{eq:weylB} and using
equation \eqref{eq:eth}
gives the radial system
\begin{equation}\label{eq:radialEF}
\Big(\frac{d}{dr}-\frac{2i\omega}{\sqrt{fg}}\Big)R_-=-\frac{\lambda}{\sqrt{gh}}\,R_+\,,\qquad
\frac{dR_+}{dr}=-\frac{\lambda}{\sqrt{gh}}\,R_-\,.
\end{equation}
For $f=g$, $h=r^2$ these reduce to the RN radial equations of \cite{Pang:2025myy}.

\subsection{The static case}
\label{sec:staticlimit}

For static case $\omega=0$, both radial coefficients in equation \eqref{eq:radialEF}
are pure $1/\sqrt{gh}$ factors, and can be absorbed by a new variable $\chi$
defined as
\begin{equation}\label{eq:chidef}
d\chi=\frac{dr}{\sqrt{g(r)h(r)}}\,,
\end{equation}
so the radial system \eqref{eq:radialEF} reduces to
\begin{equation}\label{eq:radialsystem}
\frac{dR_-}{d\chi}=-\lambda R_+\,,\qquad
\frac{dR_+}{d\chi}=-\lambda R_-\,,
\end{equation}
a constant-coefficient system for \emph{arbitrary} static spherical metrics. Each
component also satisfies the second-order equation
\begin{equation}\label{eq:secondorder}
\Big(\frac{d^2}{d\chi^2}-\lambda^2\Big)R_\pm=0\,,
\end{equation}
automatically factorized as $(\partial_\chi\mp\lambda)(\partial_\chi\pm\lambda)$.
The whole static fermionic problem is therefore controlled by a single function of
the spatial metric, the product $gh$, or equivalently the conformally invariant
``log-radius'' $\chi$ of the slice ($dl^2=dr^2/g+h\,d\Omega^2=h(d\chi^2+d\Omega^2)$).
The lapse $f$ appears nowhere. The additive constant in $\chi$ is fixed by
choosing $\chi$ to vanish on the horizon,
\begin{equation}\label{eq:chigauge}
\chi(r)=\int_{r_+}^r\frac{d\tilde{r}}{\sqrt{g(\tilde{r})h(\tilde{r})}}\,,
\end{equation}
where $r_+$ is the outermost root of $g$ (at a non-degenerate horizon the integrand
is integrable, $1/\sqrt{gh}\sim(r-r_+)^{-1/2}$, so $\chi(r_+)=0$ is finite).

\section{Exact solutions and the general TLN}
\label{sec:tln}

\subsection{Exact solutions and horizon regularity}
\label{sec:exactsol}

Equation \eqref{eq:radialsystem} has constant coefficients, so the static problem
is exactly solvable for an \emph{arbitrary} static spherical metric, the solution
being reduced to the single quadrature \eqref{eq:chigauge} defining $\chi$:
\begin{equation}\label{eq:exactsol}
R_-(r)=a\,\mathrm{e}^{\lambda\chi}+b\,\mathrm{e}^{-\lambda\chi}\,,\qquad
R_+(r)=-a\,\mathrm{e}^{\lambda\chi}+b\,\mathrm{e}^{-\lambda\chi}\,,
\end{equation}
with $a,b$ constants.

It remains to impose the physical boundary condition at the horizon. Since the EF
tetrad \eqref{eq:tetrad} is regular at the horizon, regularity of the spinor
$\psi^A$ is equivalent to regularity of its dyad components \eqref{eq:ansatz}, so
$\psi^1\propto f^{-1/2}h^{-1/2}R_-$ and $\psi^2\propto h^{-1/2}R_+$ must both be
finite at $r_+$. While $R_+$ is automatically finite, $f^{-1/2}$ diverges, so
$R_-$ must vanish at the horizon at least as fast as $\sqrt f$. Since
$\chi(r_+)=0$ is finite, this fixes $b=-a$ in equation \eqref{eq:exactsol}; choosing the
overall normalization $a=\frac12$, the unique regular solution is
\begin{equation}\label{eq:sinh}
R_-(r)=\sinh\big[\lambda\chi(r)\big]\,,\qquad
R_+(r)=-\cosh\big[\lambda\chi(r)\big]\,.
\end{equation}
More precisely, near the horizon $\chi=2\sqrt{(r-r_+)}/\sqrt{(gh)'(r_+)}+O\big[(r-r_+)^{3/2}\big]$
and $\sqrt f=\sqrt{f'(r_+)(r-r_+)}\,\big[1+O(r-r_+)\big]$, so the EF-sensitive component
approaches the finite limit
\begin{equation}\label{eq:EFlimit}
\frac{R_-}{\sqrt f}\;\longrightarrow\;
\frac{2\lambda}{\sqrt{(gh)'(r_+)\,f'(r_+)}}\,,
\end{equation}
matching the regularity conditions used in the RN analyses
\cite{Chakraborty:2025zyb,Pang:2025myy}. The finite limit requires a
non-degenerate horizon, $(gh)'(r_+)\neq0$; at a degenerate horizon
$(gh)'(r_+)=0$ and $\chi\to-\infty$, the regularity breaks down, leads to vanishing fermionic TLNs (see section~\ref{sec:vanishing}).

It is worth pointing out that evaluating the
regular solution \eqref{eq:sinh}, or the TLN \eqref{eq:tlnformula} below,
still requires the metric functions to be given explicitly and the integral
\eqref{eq:chigauge} to be evaluated, analytically or numerically, so the closed
forms do not deliver the answers without effort. Their content is structural
rather than computational, since they identify the exact object controlling the
response, the constant $\tilde\chi_0$ of equation \eqref{eq:chi0def}, and with it
the precise structure responsible for the vanishing of the fermionic TLN, which
makes the general criterion of section~\ref{sec:vanishing} possible.

\subsection{The general formula of static fermionic TLNs}
\label{sec:formula}

The falloff \eqref{eq:falloff} implies $1/\sqrt{gh}=1/r+O(1/r^2)$ at
infinity. Integrating term by term, the coordinate $\chi$ of equation
\eqref{eq:chigauge} approaches a logarithm%
\begin{equation}\label{eq:chiasym}
\chi(r) = \ln\frac{r}{r_+} + \tilde\chi_0 + \frac{c_1}{r} + \cdots,
\end{equation}
with the finite constant
\begin{equation}\label{eq:chi0def}
\tilde\chi_0 = \lim_{r\to\infty}\Big[\int_{r_+}^{r}\frac{d\tilde{r}}{\sqrt{g(\tilde{r})h(\tilde{r})}}
-\ln\frac{r}{r_+}\Big]\,.
\end{equation}
The two branches $\mathrm{e}^{\pm\lambda\chi}$ are thus the growing tidal mode
$\sim(r/r_+)^{\lambda}$ and the decaying response
$\sim(r/r_+)^{-\lambda}$ (the physical field carries an extra $h^{-1/2}\sim r^{-1}$,
giving the standard fermionic multipole powers). Expanding the regular solution
\eqref{eq:sinh} at infinity,
\begin{align}
R_-(r)&=\frac12\,\mathrm{e}^{\lambda\tilde\chi_0}\Big(\frac{r}{r_+}\Big)^{\lambda}
\big[1+O(r^{-1})\big]
-\frac12\,\mathrm{e}^{-\lambda\tilde\chi_0}\Big(\frac{r}{r_+}\Big)^{-\lambda}
\big[1+O(r^{-1})\big]\,,\label{eq:expansionm}\\
R_+(r)&=-\frac12\,\mathrm{e}^{\lambda\tilde\chi_0}\Big(\frac{r}{r_+}\Big)^{\lambda}
\big[1+O(r^{-1})\big]
-\frac12\,\mathrm{e}^{-\lambda\tilde\chi_0}\Big(\frac{r}{r_+}\Big)^{-\lambda}
\big[1+O(r^{-1})\big]\,,\label{eq:expansionp}
\end{align}
the static fermionic TLN can be read off directly as the ratio of the decaying
to the growing coefficient \cite{Chakraborty:2025zyb, Chakraborty:2026qru},
\begin{equation}\label{eq:tlnformula}
\boxed{\ \mathcal{F}_{\pm\frac12 lm} = \pm\exp\big(-2\lambda\tilde\chi_0\big)
= \pm\exp\big[-(2l+1)\tilde\chi_0\big]\,,\qquad l=\frac12,\frac32,\dots\ }
\end{equation}
independent of $m$ by spherical symmetry.

Table~\ref{tab:checks} lists the static fermionic TLNs of
equation \eqref{eq:tlnformula} for the black-hole families considered in this paper.
The Schwarzschild and RN results reproduce the known values
\cite{Chakraborty:2025zyb,Pang:2025myy}, including the vanishing extremal limit
$Q\to M$, where $\mathcal{F}_{\pm\frac12 lm}\propto(1-r_-/r_+)^{2\lambda}\to0$
\cite{Pang:2025myy}. The CSV and remnant results are new and are derived
in section~\ref{sec:realize}.

\begin{table}[h]
\centering
\resizebox{\textwidth}{!}{%
\begin{tabular}{@{}llll@{}}
\toprule
Metric & $g(r)$ & $\tilde\chi_0$ & $\mathcal{F}_{\pm\frac12 lm}$ \\
\midrule
Schwarzschild & $1-2M/r$ & $\ln 4$ & $\pm 4^{-2\lambda}$ \\
RN, charge $Q$ & $1-\frac{2M}{r}+\frac{Q^2}{r^2}$ &
$\ln\Big(\frac{2r_+}{\sqrt{M^2-Q^2}}\Big)$ &
$\pm 4^{-2\lambda}\Big(1-\frac{r_-}{r_+}\Big)^{2\lambda}$ \\
extremal RN & $(1-M/r)^2$ & $\to\infty$ & $0$ \\
\midrule
CSV \cite{Culetu:2013fsa,Simpson:2019mud}, $a<2M/\mathrm{e}$ &
$1-\frac{2M}{r}\mathrm{e}^{-a/r}$ & numerical, equation \eqref{eq:chi0def} &
$\pm \mathrm{e}^{-2\lambda\tilde\chi_0(a)}$ \\
extremal CSV, $a=2M/\mathrm{e}$ & $1-\frac{2M}{r}\mathrm{e}^{-2M/(\mathrm{e}r)}$ & $\to\infty$ & $0$ \\
LQG remnant \cite{Borges:2023fub}, $Q=0$ &
$\Big(1-\frac{2m}{r}\Big)\Big(1-\frac{r_0}{r}\Big)$ &
$\ln\Big(\frac{8m}{2m-r_0}\Big)$ &
$\pm\Big(\frac{2m-r_0}{8m}\Big)^{2\lambda}$ \\
extremal remnant, $2m=r_0$ & $(1-r_0/r)^2$ & $\to\infty$ & $0$ \\
\bottomrule
\end{tabular}}%
\caption{\label{tab:checks}Static fermionic TLNs from equation \eqref{eq:tlnformula}.
The Schwarzschild and RN results match
\cite{Chakraborty:2025zyb,Pang:2025myy}, the CSV and
LQG remnant results are new (section~\ref{sec:realize}), and $r_0=2^{-1/2}\ell_P$
for the remnant family, with $\ell_P$ the Planck length. In each family the TLN vanishes at the degenerate-horizon
configuration.}
\end{table}

\section{The vanishing of static fermionic TLNs for extremal black holes}
\label{sec:vanishing}

When $(gh)'(r_+)=0$, the regular condition \eqref{eq:EFlimit} diverges and $\chi\to-\infty$;
at such a degenerate horizon the decaying branch of equation \eqref{eq:exactsol}
diverges, regularity admits only the pure growing branch, and the fermionic TLNs
vanish.

More precisely, from equation \eqref{eq:tlnformula}, the TLN vanishes if
and only if $\tilde\chi_0\to\infty$. The integral in equation \eqref{eq:chi0def} is by itself logarithmically divergent at
infinity; since that logarithm is subtracted, the divergence must come from the
near-horizon end, where $gh$ can be parametrized as
\begin{equation}\label{eq:criterion}
gh \;\longrightarrow\; c\,(r-r_+)^{\alpha}\,,
\end{equation}
with $c>0$; the integrand of $\chi$ is then $\propto(r-r_+)^{-\alpha/2}$, whose
integral near the horizon converges for $\alpha<2$ and diverges for $\alpha\ge2$. The static
fermionic TLN therefore vanishes precisely when $gh$ has a zero of order at
least two at the horizon, generically a double zero. We assume $h$ regular
and positive at $r_+$ (a zero of $h$ there would be a collapsing two-sphere,
not a regular horizon); equation \eqref{eq:criterion} is then equivalently
$\int^{r_+}dr/\sqrt g=\infty$. The horizon lies at infinite proper distance
along the $t=\mathrm{const}$ slices, an infinite spatial throat as for
extremal RN, and the surface gravity vanishes,
$\kappa^2=\frac14 f'g'|_{r_+}=0$. These are precisely the extremal (degenerate, zero-temperature)
horizons \cite{Kunduri:2013gce}.

A related
instance of this mode exclusion is known for bosons. For extremal RN and
Kerr--Newman black holes, the static scalar TLNs vanish because a conformal
inversion isometry maps the decaying mode at infinity to a divergent mode at the
horizon \cite{Kehagias:2024yzn}. That isometry is a symmetry of the extremal
near-horizon region and therefore belongs to precisely the degenerate-horizon
class singled out by equation \eqref{eq:criterion}.

\subsection{Physical realizations: regular and remnant black holes}
\label{sec:realize}

We now exhibit the vanishing criterion at work on two spacetime families from the
literature.

\begin{figure}[t]
\centering
\includegraphics[width=0.72\textwidth]{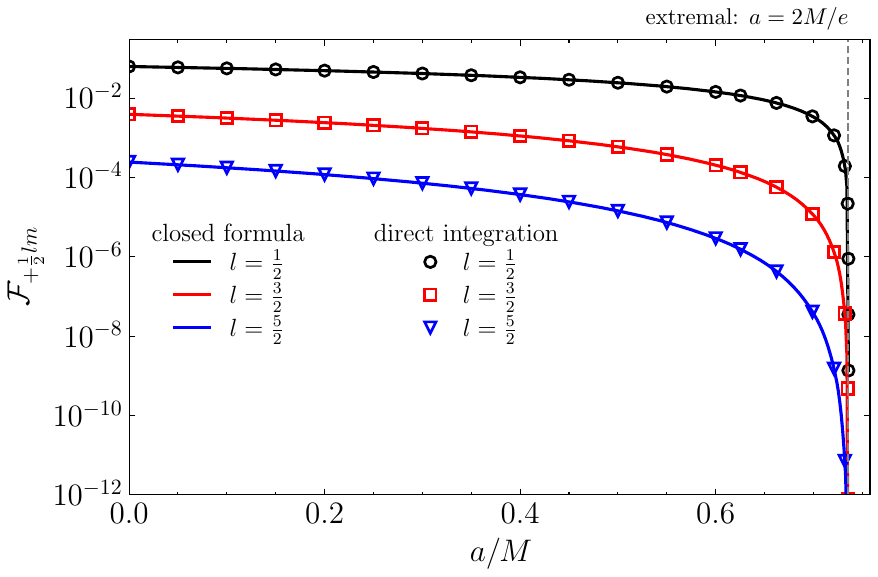}
\caption{\label{fig:sv}Static fermionic TLNs of the CSV regular black
hole \eqref{eq:sv} versus the suppression parameter $a$, for
$l=\frac12,\frac32,\frac52$. Lines show the closed formula
$\mathrm{e}^{-(2l+1)\tilde\chi_0}$ [equation \eqref{eq:tlnformula}, $\mathcal{F}_{+\frac12 lm}$
branch], and markers show direct numerical integration of the static Dirac system.
The TLNs vanish at the extremal
point $a=2M/\mathrm{e}$ (vertical dashed line), where the horizon degenerates; at $a=0$
the Schwarzschild values $4^{-(2l+1)}$ are recovered.}
\end{figure}

\paragraph{CSV regular black hole.} The Minkowski-core regular black
hole of \cite{Culetu:2013fsa,Simpson:2019mud} has
\begin{equation}\label{eq:sv}
f(r)=g(r)=1-\frac{2M}{r}\,\mathrm{e}^{-a/r}\,,\qquad h=r^2\,,
\end{equation}
with $a\ge0$ the exponential-suppression parameter ($a=0$ is Schwarzschild). The
horizons are exact in terms of the Lambert function,
$r_\pm=-a/W_{0,-1}(-a/2M)$, real for $a\le a_c\equiv2M/\mathrm{e}$, merging at
$r_c=2M/\mathrm{e}$ where the two $W$ branches coalesce. At $a=a_c$, $f(r_c)=f'(r_c)=0$
with $f''(r_c)=1/r_c^2$, so $gh=r^2f=\frac12(r-r_c)^2+O\big[(r-r_c)^3\big]$ has
a double zero, i.e.\ a degenerate horizon with an infinite throat (the geometry
stays regular, $R\to0$ at the core), and the fermionic TLN vanishes. Near
extremality $r_+-r_-\simeq2\sqrt{2a_c(a_c-a)}$ and
$\tilde\chi_0\sim\sqrt2\,\ln\frac{r_c}{r_+-r_-}$ (using $r_c\sqrt{f''(r_c)}=1$),
so
\begin{equation}\label{eq:svpower}
\mathcal{F}_{\pm\frac12 lm}\propto \big(1-a/a_c\big)^{\sqrt2\,\lambda}
\qquad(a\to a_c^-)\,.
\end{equation}
Figure~\ref{fig:sv} shows $\mathcal{F}_{+\frac12 lm}$ as a function of $a$ for
$l=\frac12,\frac32,\frac52$, clearly displaying the vanishing of the TLNs as $a$
approaches the extremal value $a_c$. The
Bardeen and Hayward regular black holes \cite{bardeen1968,Hayward:2005gi} behave
identically in this respect. Their fermionic TLNs vanish at the extremal
configuration and are nonzero elsewhere\footnote{It's worth pointing out that our criterion only works for outer horizons; the inner degenerate horizons of regular black holes, like those in \cite{Feng:2026sra}, do not result in vanishing fermionic TLNs.}.

\begin{figure}[t]
\centering
\includegraphics[width=0.72\textwidth]{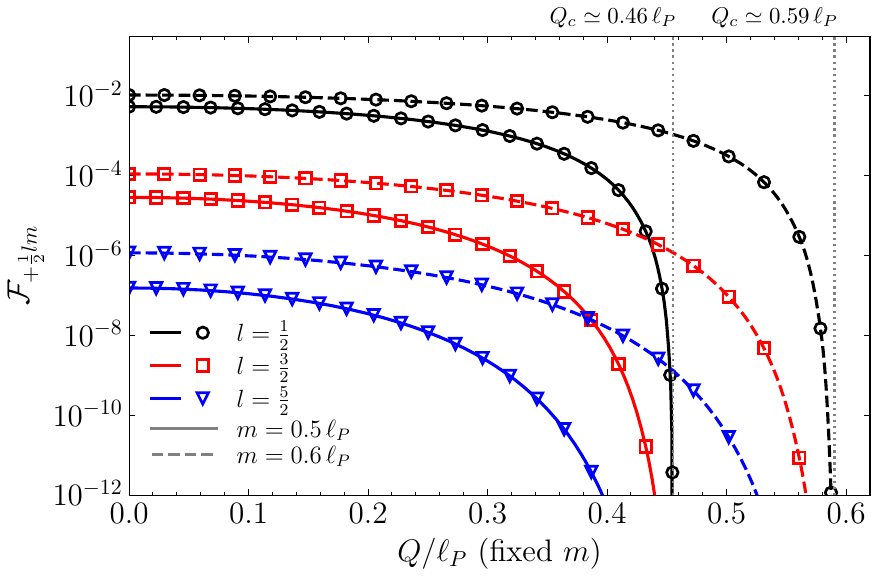}
\caption{\label{fig:remnant}Static fermionic TLNs of the LQG remnant black-hole
family \cite{Borges:2023fub} versus the charge $Q$ at fixed masses
$m=0.5\,\ell_P$ (solid) and $m=0.6\,\ell_P$ (dashed), for
$l=\frac12,\frac32,\frac52$. Lines show the closed formula \eqref{eq:tlnformula}
($\mathcal{F}_{+\frac12 lm}$ branch), and markers show direct numerical
integration. The TLNs vanish as $Q$ reaches the charged-remnant value
$Q_c(m)=\sqrt{\sqrt2\,m\ell_P-\ell_P^2/2}$ (dotted vertical lines), explicitly
$Q_c\simeq0.46\,\ell_P$ for $m=0.5\,\ell_P$ and $Q_c\simeq0.59\,\ell_P$ for
$m=0.6\,\ell_P$, where the horizons merge with the transition surface.}
\end{figure}

\paragraph{LQG remnant black holes.} The LQG remnant black-hole family
of Borges et al.\ \cite{Borges:2023fub}, with the minimal-area condition
imposed on the transition surface, has $f=1-2m/r+Q^2/r^2$ and\footnote{For $Q=0$
this reduces to $gh=(r-2m)(r-r_0)$ with $r_0=\ell_P/\sqrt2$ fixed for all $m$,
whence the closed form of table~\ref{tab:checks}.}
\begin{equation}\label{eq:remgh}
gh=\frac{(r-r_h^+)(r-r_h^-)(r-r_0)(r-r_2)}{r^2}\,,\qquad
r_2=\frac{Q^2}{m(1+\mathfrak{s})}\,,\;\;
\mathfrak{s}=\sqrt{1-\frac{b_0^2Q^2}{(b_0^2-1)m^2}}\,,
\end{equation}
with $r_h^\pm=m\pm\sqrt{m^2-Q^2}$ the classical radii, $r_0$ the black-to-white
transition radius, and $b_0^2=1+\delta_b^2/12$ with
$\delta_b^2=12\ell_P^2/(2\sqrt2\,m\ell_P-2Q^2-\ell_P^2)$, with $\ell_P$ the
Planck length. The remnant states, the endpoint of
Hawking evaporation in this model, sit on the curve
\begin{equation}\label{eq:remnantcurve}
2\sqrt2\,m\ell_P=2Q^2+\ell_P^2\qquad(\delta_b\to\infty)\,,
\end{equation}
along which $r_0\to r_h^+$ and $r_2\to r_h^-$, so
$gh\to(r-r_h^+)^2(r-r_h^-)^2/r^2$, a double zero at the outer horizon for every
allowed charge $0\le Q\le\ell_P/\sqrt2$. Note that the lapse keeps a simple zero
there, $f'(r_h^+)\neq0$ for $Q<\ell_P/\sqrt2$; the degeneracy resides entirely in
the double zero of $g$, so these remnant horizons are degenerate without a double
zero of the lapse. The fermionic TLN therefore vanishes at
the extremal remnant for every $Q$ (for $Q=0$ at $m=\ell_P/(2\sqrt2)$) and nowhere
off the curve (figure~\ref{fig:remnant}).

Finally, note that the scalar TLNs of these two families cannot vanish. The
vanishing of the static scalar response requires the scalar ladder symmetry,
which exists only on the conformal-RN family
\eqref{eq:scalargfamily} \cite{Sharma:2024hlz,Sharma:2025xii}\footnote{The
SV metric \eqref{eq:sv} contains the exponential $\mathrm{e}^{-a/r}$, and the
remnant metric has $f\neq g$; neither is of the form \eqref{eq:scalargfamily}.},
and neither the CSV nor the LQG remnant black holes belong to it\footnote{We have also verified numerically by direct integration that the scalar TLNs are indeed nonzero even in the extremal case.}. These are thus two explicit examples of black
holes with vanishing fermionic TLNs but nonvanishing bosonic ones. For
completeness, the static scalar equation on the general metric is derived in the
NP formalism in appendix~\ref{app:scalar}.

\section{Conclusions}
\label{sec:conclusions}

We have analyzed the static tidal response of a massless Dirac field on an
arbitrary static, spherically symmetric black hole. Derived in the NP formalism
in an EF tetrad, the static Dirac equation reduces to a
constant-coefficient system in the conformal coordinate $\chi$ of the spatial
slices and is exactly solvable for arbitrary metrics, with the lapse $f$ dropping
out entirely. Horizon regularity then yields a closed-form expression for the
static fermionic TLN of any asymptotically flat black hole, which
reproduces the known Schwarzschild and RN values. The response is therefore
controlled by a single conformal invariant of the spatial geometry. In contrast
to bosons, whose TLNs vanish for all Schwarzschild and Kerr black holes,
fermionic TLNs are generically nonzero, and they vanish if and only if $gh$ has
at least a double zero at the horizon, i.e.\ at an extremal horizon
with an infinite spatial throat.

Two physical families realize the vanishing class. The fermionic TLNs of the CSV
regular black hole vanish at its extremal configuration, and those of the LQG
remnant black holes \cite{Borges:2023fub} vanish at the extremal remnant,
including along the whole charged remnant curve, with the neutral case given in
closed form. The static scalar response can vanish only on the
conformal-RN family \eqref{eq:scalargfamily}
\cite{Sharma:2024hlz,Sharma:2025xii}, and neither the CSV nor the LQG
remnant black holes belong to it. These are therefore
explicit examples of black holes with vanishing fermionic but nonvanishing
bosonic TLNs. This difference between the two sectors traces back to the order
of the wave operator. The static, first-order Dirac operator reads only the
conformal geometry of the spatial slice and is blind to the lapse $f$. The
second-order bosonic operator feels the lapse explicitly, and this is why the
ladder symmetry constrains the background in the bosonic case but not in the
fermionic one.

Several extensions are worth pursuing. The most pressing is the rotating case.
For a Kerr black hole the static fermionic response is finite and nonzero even at
extremality \cite{Chakraborty:2025zyb}, in striking contrast to the spherical
criterion established here, although the extremal Kerr horizon also lies at
infinite proper distance. Understanding what replaces the infinite-throat
criterion under rotation, and how the radial and angular sectors share the
response, would clarify the geometric content of the vanishing. A second direction
is the response at nonzero frequency, where
the lapse $f$ re-enters the radial system and the response acquires an absorptive
part. This would connect the present framework with the dynamical tidal response
\cite{Chakraborty:2023zed,Perry:2023wmm,Chakraborty:2025wvs,HegadeKR:2025qwj,Chakraborty:2026dox}
and with the peculiar stability properties of extremal horizons, on which the
static vanishing may shed some light. Finally, it would be interesting to relate
$\tilde\chi_0$ more directly to the near-horizon geometry, for instance to the
AdS$_2$ throat that underlies the bosonic vanishing
\cite{Charalambous:2021mea,Berens:2025okm}.

\acknowledgments

This work is partially supported by the National Key Research and Development
Program of China with Grant No.~2021YFC2203001 as well as the National Natural
Science Foundation of China (NSFC) with Grant Nos.~12035016, 12275350, 12375048,
12375058, 12361141825, 12447182, 12575047, and 12505082. X.~P. is also supported
by the Doctoral Initiation Grant 24KE051 and Basic Research Grant 25kx010 from
China West Normal University.

\appendix

\section{The scalar equation in the NP formalism}
\label{app:scalar}

For comparison with the fermionic problem of section~\ref{sec:dirac}, we derive the
massless scalar (Klein--Gordon) equation $\Box\Phi=0$ on the EF metric
\eqref{eq:metricEF} in the NP formalism, using the same regular tetrad
\eqref{eq:tetrad}. For a scalar,
$\Box\Phi=g^{\mu\nu}\nabla_\mu\nabla_\nu\Phi$ with
$g^{\mu\nu}=2l^{(\mu}n^{\nu)}-2m^{(\mu}\bar m^{\nu)}$; writing this through the NP
operators produces the two commutator vectors
$A^\nu\equiv l^\mu\nabla_\mu n^\nu+n^\mu\nabla_\mu l^\nu$ and
$B^\nu\equiv m^\mu\nabla_\mu\bar m^\nu+\bar m^\mu\nabla_\mu m^\nu$, which for the
tetrad \eqref{eq:tetrad} evaluate to
\begin{equation}\label{eq:commvec}
A^\nu=2\varepsilon\,n^\nu\,,\qquad
B^\nu=-2\mu\,l^\nu+2\rho\,n^\nu-2\alpha\,(m^\nu+\bar m^\nu)\,.
\end{equation}
The scalar wave operator in NP form is therefore
\begin{equation}\label{eq:boxNP}
\Box=D\Delta+\Delta D-\delta\bar\delta-\bar\delta\delta
-2\varepsilon\,\Delta-2\mu\,D+2\rho\,\Delta-2\alpha\,(\delta+\bar\delta)\,,
\end{equation}
which coincides with the Laplace--Beltrami operator for arbitrary $f,g,h$.
Substituting the tetrad explicitly,
\begin{equation}\label{eq:scalarPDE}
\partial_r\big(\sqrt{fg}\,h\,\partial_r\Phi\big)+2h\,\partial_v\partial_r\Phi
+h'\,\partial_v\Phi+\sqrt{\frac fg}\,L^2_{S^2}\Phi=0\,,
\end{equation}
with $L^2_{S^2}$ the unit-sphere Laplacian. Separating
$\Phi=\mathrm{e}^{-i\omega v}\phi_\ell(r)Y_{\ell m}(\theta,\varphi)$ and passing to the
coordinate $\chi$ of equation \eqref{eq:chidef}, using
$\sqrt{fg}\,h/\sqrt{gh}=\sqrt{fh}$, gives
\begin{equation}\label{eq:radScalChi}
\partial_\chi\big(\sqrt{fh}\,\partial_\chi\phi_\ell\big)
-2i\omega h\,\partial_\chi\phi_\ell-i\omega\,(\partial_\chi h)\,\phi_\ell
-\ell(\ell+1)\sqrt{fh}\,\phi_\ell=0\,,
\end{equation}
and in the static limit
\begin{equation}\label{eq:statScalChi}
\partial_\chi\big(\sqrt{fh}\,\partial_\chi\phi_\ell\big)
-\ell(\ell+1)\sqrt{fh}\,\phi_\ell=0
\quad\Longleftrightarrow\quad
\phi_\ell''+\partial_\chi\!\big(\ln\sqrt{fh}\big)\phi_\ell'-\ell(\ell+1)\phi_\ell=0\,,
\end{equation}
with primes $d/d\chi$; equivalently, setting $\phi_\ell=(fh)^{-1/4}R_\ell$ gives
the Schr\"odinger form
\begin{equation}\label{eq:scalCan}
R_\ell''-\Big[\ell(\ell+1)+V(\chi)\Big]R_\ell=0\,,\quad
V(\chi)=\frac12\frac{s''}{s}-\frac14\Big(\frac{s'}{s}\Big)^{\!2}\,,\;
s\equiv\sqrt{fh}\,.
\end{equation}
(In the $r$ coordinate, \eqref{eq:statScalChi} reads
$\partial_r\big(h\sqrt{fg}\,\partial_r\phi_\ell\big)-\ell(\ell+1)\sqrt{f/g}\,\phi_\ell=0$.)
On Schwarzschild, $\chi=\mathrm{arcosh}((r-M)/M)$,
$s=M\sinh\chi$, and \eqref{eq:statScalChi} is the Legendre equation in $\cosh\chi$.

The contrast with the fermionic system \eqref{eq:radialsystem}--\eqref{eq:secondorder}
is sharp. There, the spin-connection terms are pure gradients,
$\sqrt{fg}\,\partial_r\ln(fh)^{1/4}$ in the first Weyl equation and
$\sqrt{g/f}\,\partial_r\ln h^{1/2}$ in the second, and are absorbed by the
prefactors of the ansatz \eqref{eq:ansatz}; the
remaining coefficients involve only $1/\sqrt{gh}$; the second-order form
\eqref{eq:secondorder} is \emph{free} in $\chi$-space for arbitrary metrics. The scalar
equation \eqref{eq:statScalChi} instead feels the lapse through $\sqrt{fh}$, and no
field redefinition makes it depend on $gh$ alone.

\section{The fermionic ladder for arbitrary metrics}
\label{app:ladder}

The vanishing of the bosonic TLNs of Schwarzschild and RN black holes is tied to a
ladder symmetry of the static perturbation equations, which for scalars exists only
on the conformal-RN family \eqref{eq:scalargfamily}
\cite{Sharma:2024hlz,Sharma:2025xii}. Here we derive the corresponding
fermionic ladder and show that it exists for \emph{arbitrary} metrics.

Write the first-order system \eqref{eq:radialsystem} as
\begin{equation}\label{eq:Hlam}
\mathcal H_\lambda\begin{pmatrix}R_-\\R_+\end{pmatrix}=0\,,\qquad
\mathcal H_\lambda\equiv\frac{d}{d\chi}-\lambda M\,,\qquad M=-\sigma_1\,,\quad
M^2=\mathbf 1\,,
\end{equation}
and seek raising operators $D^+_\lambda=a(\chi)\,d/d\chi+b(\chi)+\lambda c(\chi)$
with $\lambda$-independent matrix coefficients, intertwining as
$\mathcal H_{\lambda+1}D^+_\lambda=D^+_\lambda\mathcal H_\lambda$. Applied to an
arbitrary spinor, the residual is
\begin{equation}\label{eq:residual}
\mathcal H_{\lambda+1}D^+_\lambda-D^+_\lambda\mathcal H_\lambda
= \big(a'-Ma+\lambda[a,M]\big)\partial_\chi
+\big(b'+\lambda c'-M(b+\lambda c)+\lambda[b+\lambda c,M]\big)\,.
\end{equation}
Requiring the residual to vanish for arbitrary spinors and arbitrary $\lambda$
gives $[a,M]=[b,M]=[c,M]=0$ and
$a'=Ma$, $b'=Mb$, $c'=Mc$, hence
\begin{equation}\label{eq:laddersol}
D^+_\lambda=\mathrm{e}^{M\chi}\Big(a_0\frac{d}{d\chi}+b_0+\lambda c_0\Big)\,,
\end{equation}
with $a_0,b_0,c_0$ arbitrary constant matrices commuting with $M$ (for $M=-\sigma_1$,
elements of $\mathrm{span}\{\mathbf 1,M\}$); since $M$ commutes with its own
exponential, substituting back into equation \eqref{eq:residual} the residual vanishes
identically. Similarly $D^-_\lambda=\mathrm{e}^{-M\chi}(a_0\,d/d\chi+b_0+\lambda c_0)$
realizes $\mathcal H_{\lambda-1}D^-_\lambda=D^-_\lambda\mathcal H_\lambda$. For
$M^2=\mathbf 1$ the exponential reduces to $\mathrm{e}^{\pm M\chi}=\cosh\chi\,\mathbf
1\pm M\sinh\chi$, with eigenvalues $\mathrm{e}^{\pm\chi}$. The minimal choice
$a_0=c_0=0$, $b_0=\mathbf 1$ gives the pure multiplication ladder
$D^\pm=\mathrm{e}^{\pm M\chi}$, which acts diagonally on the two eigencomponents
of $M$, multiplying the $\mathrm{e}^{\lambda\chi}$ branch of a solution by
$\mathrm{e}^{\chi}$ and the $\mathrm{e}^{-\lambda\chi}$ branch by
$\mathrm{e}^{-\chi}$. Any $\lambda$-solution is thus mapped to a
$(\lambda+1)$-solution with the same constants $a,b$, and $D^-$ analogously to a
$(\lambda-1)$-solution.

The point of this derivation is what does \emph{not} appear, namely any condition
on the metric. Because the static fermionic problem has constant coefficients in
$\chi$ for arbitrary $(f,g,h)$, the intertwining equations determine the ladder
coefficients completely while leaving the background arbitrary. This is in sharp
contrast to the scalar case, where the ladder exists only if the background is
constrained to the conformal-RN family \eqref{eq:scalargfamily}
\cite{Sharma:2024hlz}. Note
also that the fermionic ladder is intrinsically first order. A direct second-order
ansatz $D^+_\lambda=-A(\chi)\partial_\chi+(\lambda+1)B(\chi)$ acting on the operator in
\eqref{eq:secondorder} has only the trivial solution, because the eigenvalue step
$\lambda^2\to(\lambda+1)^2$ is $2\lambda+1$ and never equals the $2(\lambda+1)$
required by the ansatz coefficient, unlike the bosonic Casimir $\ell(\ell+1)$, whose
step is exactly $2(\ell+1)$ and therefore admits the scalar ladder.

\providecommand{\href}[2]{#2}\begingroup\raggedright\endgroup


\begin{thebibliography}{10}

\bibitem{Love1909}
A.E.H.~Love, \emph{The yielding of the {Earth} to disturbing forces}, {\emph{Proc. R. Soc. Lond. A} {\bfseries 82} (1909) 73}.

\bibitem{Goldberger:2004jt}
W.D.~Goldberger and I.Z.~Rothstein, \emph{{An Effective field theory of gravity for extended objects}}, \href{https://doi.org/10.1103/PhysRevD.73.104029}{\emph{Phys. Rev. D} {\bfseries 73} (2006) 104029} [\href{https://arxiv.org/abs/hep-th/0409156}{{\ttfamily hep-th/0409156}}].

\bibitem{Damour:2009va}
T.~Damour and O.M.~Lecian, \emph{{On the gravitational polarizability of black holes}}, \href{https://doi.org/10.1103/PhysRevD.80.044017}{\emph{Phys. Rev. D} {\bfseries 80} (2009) 044017} [\href{https://arxiv.org/abs/0906.3003}{{\ttfamily 0906.3003}}].

\bibitem{Binnington:2009bb}
T.~Binnington and E.~Poisson, \emph{{Relativistic theory of tidal Love numbers}}, \href{https://doi.org/10.1103/PhysRevD.80.084018}{\emph{Phys. Rev. D} {\bfseries 80} (2009) 084018} [\href{https://arxiv.org/abs/0906.1366}{{\ttfamily 0906.1366}}].

\bibitem{Damour:2009vw}
T.~Damour and A.~Nagar, \emph{{Relativistic tidal properties of neutron stars}}, \href{https://doi.org/10.1103/PhysRevD.80.084035}{\emph{Phys. Rev. D} {\bfseries 80} (2009) 084035} [\href{https://arxiv.org/abs/0906.0096}{{\ttfamily 0906.0096}}].

\bibitem{Fang:2005qq}
H.~Fang and G.~Lovelace, \emph{{Tidal coupling of a Schwarzschild black hole and circularly orbiting moon}}, \href{https://doi.org/10.1103/PhysRevD.72.124016}{\emph{Phys. Rev. D} {\bfseries 72} (2005) 124016} [\href{https://arxiv.org/abs/gr-qc/0505156}{{\ttfamily gr-qc/0505156}}].

\bibitem{Bini:2012gu}
D.~Bini, T.~Damour and G.~Faye, \emph{{Effective action approach to higher-order relativistic tidal interactions in binary systems and their effective one body description}}, \href{https://doi.org/10.1103/PhysRevD.85.124034}{\emph{Phys. Rev. D} {\bfseries 85} (2012) 124034} [\href{https://arxiv.org/abs/1202.3565}{{\ttfamily 1202.3565}}].

\bibitem{Porto:2016pyg}
R.A.~Porto, \emph{{The effective field theorist{\textquoteright}s approach to gravitational dynamics}}, \href{https://doi.org/10.1016/j.physrep.2016.04.003}{\emph{Phys. Rept.} {\bfseries 633} (2016) 1} [\href{https://arxiv.org/abs/1601.04914}{{\ttfamily 1601.04914}}].

\bibitem{Chakraborty:2026qru}
S.~Chakraborty and P.~Pani, \emph{{Tidal Response of Compact Objects}},  \href{https://arxiv.org/abs/2604.08679}{{\ttfamily 2604.08679}}.

\bibitem{Hinderer:2007mb}
T.~Hinderer, \emph{{Tidal Love numbers of neutron stars}}, \href{https://doi.org/10.1086/533487}{\emph{Astrophys. J.} {\bfseries 677} (2008) 1216} [\href{https://arxiv.org/abs/0711.2420}{{\ttfamily 0711.2420}}].

\bibitem{Flanagan:2007ix}
E.E.~Flanagan and T.~Hinderer, \emph{{Constraining neutron star tidal Love numbers with gravitational wave detectors}}, \href{https://doi.org/10.1103/PhysRevD.77.021502}{\emph{Phys. Rev. D} {\bfseries 77} (2008) 021502} [\href{https://arxiv.org/abs/0709.1915}{{\ttfamily 0709.1915}}].

\bibitem{Damour:2012yf}
T.~Damour, A.~Nagar and L.~Villain, \emph{{Measurability of the tidal polarizability of neutron stars in late-inspiral gravitational-wave signals}}, \href{https://doi.org/10.1103/PhysRevD.85.123007}{\emph{Phys. Rev. D} {\bfseries 85} (2012) 123007} [\href{https://arxiv.org/abs/1203.4352}{{\ttfamily 1203.4352}}].

\bibitem{Yagi:2013bca}
K.~Yagi and N.~Yunes, \emph{{I-Love-Q}}, \href{https://doi.org/10.1126/science.1236462}{\emph{Science} {\bfseries 341} (2013) 365} [\href{https://arxiv.org/abs/1302.4499}{{\ttfamily 1302.4499}}].

\bibitem{Yagi:2016ejg}
K.~Yagi and N.~Yunes, \emph{{I-Love-Q Relations: From Compact Stars to Black Holes}}, \href{https://doi.org/10.1088/0264-9381/33/9/095005}{\emph{Class. Quant. Grav.} {\bfseries 33} (2016) 095005} [\href{https://arxiv.org/abs/1601.02171}{{\ttfamily 1601.02171}}].

\bibitem{LIGOScientific:2018cki}
{\scshape LIGO Scientific, Virgo} collaboration, \emph{{GW170817: Measurements of neutron star radii and equation of state}}, \href{https://doi.org/10.1103/PhysRevLett.121.161101}{\emph{Phys. Rev. Lett.} {\bfseries 121} (2018) 161101} [\href{https://arxiv.org/abs/1805.11581}{{\ttfamily 1805.11581}}].

\bibitem{Gurlebeck:2015xpa}
N.~G{\"u}rlebeck, \emph{{No-hair theorem for Black Holes in Astrophysical Environments}}, \href{https://doi.org/10.1103/PhysRevLett.114.151102}{\emph{Phys. Rev. Lett.} {\bfseries 114} (2015) 151102} [\href{https://arxiv.org/abs/1503.03240}{{\ttfamily 1503.03240}}].

\bibitem{LeTiec:2020bos}
A.~Le~Tiec, M.~Casals and E.~Franzin, \emph{{Tidal Love Numbers of Kerr Black Holes}}, \href{https://doi.org/10.1103/PhysRevD.103.084021}{\emph{Phys. Rev. D} {\bfseries 103} (2021) 084021} [\href{https://arxiv.org/abs/2010.15795}{{\ttfamily 2010.15795}}].

\bibitem{Chia:2020yla}
H.S.~Chia, \emph{{Tidal deformation and dissipation of rotating black holes}}, \href{https://doi.org/10.1103/PhysRevD.104.024013}{\emph{Phys. Rev. D} {\bfseries 104} (2021) 024013} [\href{https://arxiv.org/abs/2010.07300}{{\ttfamily 2010.07300}}].

\bibitem{Poisson:2020mdi}
E.~Poisson, \emph{{Gravitomagnetic Love tensor of a slowly rotating body: post-Newtonian theory}}, \href{https://doi.org/10.1103/PhysRevD.102.064059}{\emph{Phys. Rev. D} {\bfseries 102} (2020) 064059} [\href{https://arxiv.org/abs/2007.01678}{{\ttfamily 2007.01678}}].

\bibitem{Hui:2020xxx}
L.~Hui, A.~Joyce, R.~Penco, L.~Santoni and A.R.~Solomon, \emph{{Static response and Love numbers of Schwarzschild black holes}}, \href{https://doi.org/10.1088/1475-7516/2021/04/052}{\emph{JCAP} {\bfseries 04} (2021) 052} [\href{https://arxiv.org/abs/2010.00593}{{\ttfamily 2010.00593}}].

\bibitem{Hui:2021vcv}
L.~Hui, A.~Joyce, R.~Penco, L.~Santoni and A.R.~Solomon, \emph{{Ladder symmetries of black holes. Implications for love numbers and no-hair theorems}}, \href{https://doi.org/10.1088/1475-7516/2022/01/032}{\emph{JCAP} {\bfseries 01} (2022) 032} [\href{https://arxiv.org/abs/2105.01069}{{\ttfamily 2105.01069}}].

\bibitem{Berens:2022ebl}
R.~Berens, L.~Hui and Z.~Sun, \emph{{Ladder symmetries of black holes and de Sitter space: love numbers and quasinormal modes}}, \href{https://doi.org/10.1088/1475-7516/2023/06/056}{\emph{JCAP} {\bfseries 06} (2023) 056} [\href{https://arxiv.org/abs/2212.09367}{{\ttfamily 2212.09367}}].

\bibitem{Charalambous:2021mea}
P.~Charalambous, S.~Dubovsky and M.M.~Ivanov, \emph{{On the Vanishing of Love Numbers for Kerr Black Holes}}, \href{https://doi.org/10.1007/JHEP05(2021)038}{\emph{JHEP} {\bfseries 05} (2021) 038} [\href{https://arxiv.org/abs/2102.08917}{{\ttfamily 2102.08917}}].

\bibitem{Ivanov:2022qqt}
M.M.~Ivanov and Z.~Zhou, \emph{{Vanishing of Black Hole Tidal Love Numbers from Scattering Amplitudes}}, \href{https://doi.org/10.1103/PhysRevLett.130.091403}{\emph{Phys. Rev. Lett.} {\bfseries 130} (2023) 091403} [\href{https://arxiv.org/abs/2209.14324}{{\ttfamily 2209.14324}}].

\bibitem{Berens:2025okm}
R.~Berens, L.~Hui, D.~McLoughlin, R.~Penco and J.~Staunton, \emph{{Geometric symmetries for the vanishing of the black hole tidal Love numbers}}, \href{https://doi.org/10.1088/1475-7516/2026/05/096}{\emph{JCAP} {\bfseries 05} (2026) 096} [\href{https://arxiv.org/abs/2510.18952}{{\ttfamily 2510.18952}}].

\bibitem{Sharma:2024hlz}
C.~Sharma, R.~Ghosh and S.~Sarkar, \emph{{Exploring ladder symmetry and Love numbers for static and rotating black holes}}, \href{https://doi.org/10.1103/PhysRevD.109.L041505}{\emph{Phys. Rev. D} {\bfseries 109} (2024) L041505} [\href{https://arxiv.org/abs/2401.00703}{{\ttfamily 2401.00703}}].

\bibitem{Sharma:2025xii}
C.~Sharma, S.~Roy and S.~Sarkar, \emph{{Ladder symmetry: The necessary and sufficient condition for vanishing Love numbers}}, \href{https://doi.org/10.1103/44dg-smt2}{\emph{Phys. Rev. D} {\bfseries 113} (2026) 024066} [\href{https://arxiv.org/abs/2511.09670}{{\ttfamily 2511.09670}}].

\bibitem{Chakraborty:2023zed}
S.~Chakraborty, E.~Maggio, M.~Silvestrini and P.~Pani, \emph{{Dynamical tidal Love numbers of Kerr-like compact objects}}, \href{https://doi.org/10.1103/PhysRevD.110.084042}{\emph{Phys. Rev. D} {\bfseries 110} (2024) 084042} [\href{https://arxiv.org/abs/2310.06023}{{\ttfamily 2310.06023}}].

\bibitem{Perry:2023wmm}
M.~Perry and M.J.~Rodriguez, \emph{{Dynamical Love Numbers for Kerr Black Holes}},  \href{https://arxiv.org/abs/2310.03660}{{\ttfamily 2310.03660}}.

\bibitem{Chakraborty:2025wvs}
S.~Chakraborty, V.~De~Luca, L.~Gualtieri and P.~Pani, \emph{{Dynamical Love numbers of black holes: Theory and gravitational waveforms}}, \href{https://doi.org/10.1103/fr3y-s1sz}{\emph{Phys. Rev. D} {\bfseries 112} (2025) 104015} [\href{https://arxiv.org/abs/2507.22994}{{\ttfamily 2507.22994}}].

\bibitem{HegadeKR:2025qwj}
A.~Hegade K.~R., K.J.~Kwon, T.~Venumadhav, H.~Yu and N.~Yunes, \emph{{Relativistic and Dynamical Love Numbers}}, \href{https://doi.org/10.1103/1wdp-6x27}{\emph{Phys. Rev. Lett.} {\bfseries 136} (2026) 071401} [\href{https://arxiv.org/abs/2507.10693}{{\ttfamily 2507.10693}}].

\bibitem{Chakraborty:2026dox}
S.~Chakraborty, M.V.S.~Saketh, T.~Hinderer and J.~Steinhoff, \emph{{Dynamical tidal Love numbers of black holes under generic perturbations: Connecting black hole perturbation theory with effective field theory}},  \href{https://arxiv.org/abs/2605.00693}{{\ttfamily 2605.00693}}.

\bibitem{Apostolidis:2026qsg}
T.~Apostolidis, V.~De~Luca, L.~Gualtieri, T.~Katagiri, P.~Pani and L.~Santoni, \emph{{Dynamical Tidal Response of Neutron Stars: from Effective Field Theory to Gravitational Waveforms}},  \href{https://arxiv.org/abs/2606.19446}{{\ttfamily 2606.19446}}.

\bibitem{Kol:2011vg}
B.~Kol and M.~Smolkin, \emph{{Black hole stereotyping: Induced gravito-static polarization}}, \href{https://doi.org/10.1007/JHEP02(2012)010}{\emph{JHEP} {\bfseries 02} (2012) 010} [\href{https://arxiv.org/abs/1110.3764}{{\ttfamily 1110.3764}}].

\bibitem{Cardoso:2019vof}
V.~Cardoso, L.~Gualtieri and C.J.~Moore, \emph{{Gravitational waves and higher dimensions: Love numbers and Kaluza-Klein excitations}}, \href{https://doi.org/10.1103/PhysRevD.100.124037}{\emph{Phys. Rev. D} {\bfseries 100} (2019) 124037} [\href{https://arxiv.org/abs/1910.09557}{{\ttfamily 1910.09557}}].

\bibitem{Pang:2026vah}
X.~Pang, Y.~Tian, H.~Zhang and Q.~Jiang, \emph{{Fermionic Love number of higher-dimensional Reissner-Nordstr{\"o}m black holes}},  \href{https://arxiv.org/abs/2606.24365}{{\ttfamily 2606.24365}}.

\bibitem{Franzin:2024cah}
E.~Franzin, A.M.~Frassino and J.V.~Rocha, \emph{{Tidal Love numbers of static black holes in anti-de Sitter}}, \href{https://doi.org/10.1007/JHEP12(2024)224}{\emph{JHEP} {\bfseries 12} (2025) 224} [\href{https://arxiv.org/abs/2410.23545}{{\ttfamily 2410.23545}}].

\bibitem{Pereniguez:2025jxq}
D.~Pere{\~n}iguez and E.~Karnickis, \emph{{Nonzero Love numbers of magnetic black holes}}, \href{https://doi.org/10.1103/m8wj-gxss}{\emph{Phys. Rev. D} {\bfseries 113} (2026) L081502} [\href{https://arxiv.org/abs/2509.12418}{{\ttfamily 2509.12418}}].

\bibitem{Cardoso:2017cfl}
V.~Cardoso, E.~Franzin, A.~Maselli, P.~Pani and G.~Raposo, \emph{{Testing strong-field gravity with tidal Love numbers}}, \href{https://doi.org/10.1103/PhysRevD.95.084014}{\emph{Phys. Rev. D} {\bfseries 95} (2017) 084014} [\href{https://arxiv.org/abs/1701.01116}{{\ttfamily 1701.01116}}].

\bibitem{Cardoso:2019rvt}
V.~Cardoso and P.~Pani, \emph{{Testing the nature of dark compact objects: a status report}}, \href{https://doi.org/10.1007/s41114-019-0020-4}{\emph{Living Rev. Rel.} {\bfseries 22} (2019) 4} [\href{https://arxiv.org/abs/1904.05363}{{\ttfamily 1904.05363}}].

\bibitem{Maggio:2021ans}
E.~Maggio, P.~Pani and G.~Raposo, \emph{{Testing the nature of dark compact objects with gravitational waves}},  \href{https://arxiv.org/abs/2105.06410}{{\ttfamily 2105.06410}}.

\bibitem{DeLuca:2022tkm}
V.~De~Luca, J.~Khoury and S.S.C.~Wong, \emph{{Implications of the weak gravity conjecture for tidal Love numbers of black holes}}, \href{https://doi.org/10.1103/PhysRevD.108.044066}{\emph{Phys. Rev. D} {\bfseries 108} (2023) 044066} [\href{https://arxiv.org/abs/2211.14325}{{\ttfamily 2211.14325}}].

\bibitem{Barbosa:2025uau}
S.~Barbosa, P.~Brax, S.~Fichet and L.~de~Souza, \emph{{Running Love numbers and the Effective Field Theory of gravity}}, \href{https://doi.org/10.1088/1475-7516/2025/07/071}{\emph{JCAP} {\bfseries 07} (2025) 071} [\href{https://arxiv.org/abs/2501.18684}{{\ttfamily 2501.18684}}].

\bibitem{Motaharfar:2025typ}
M.~Motaharfar and P.~Singh, \emph{{Loop quantum gravitational signatures via Love numbers}}, \href{https://doi.org/10.1103/PhysRevD.111.106018}{\emph{Phys. Rev. D} {\bfseries 111} (2025) 106018} [\href{https://arxiv.org/abs/2501.09151}{{\ttfamily 2501.09151}}].

\bibitem{Barbosa:2026qcv}
S.~Barbosa, S.~Fichet and L.~de~Souza, \emph{{Running Love Numbers of Charged Black Holes}},  \href{https://arxiv.org/abs/2602.00349}{{\ttfamily 2602.00349}}.

\bibitem{Chakraborty:2025zyb}
S.~Chakraborty, P.~Heidmann and P.~Pani, \emph{{Fermionic response of black holes in general relativity}}, \href{https://doi.org/10.1103/2yr1-9ymw}{\emph{Phys. Rev. D} {\bfseries 113} (2026) L061503} [\href{https://arxiv.org/abs/2508.20155}{{\ttfamily 2508.20155}}].

\bibitem{Pang:2025myy}
X.~Pang, Y.~Tian, H.~Zhang and Q.~Jiang, \emph{{Fermionic Love number of Reissner-Nordstr{\"o}m black holes}}, \href{https://doi.org/10.1016/j.physletb.2026.140555}{\emph{Phys. Lett. B} {\bfseries 878} (2026) 140555} [\href{https://arxiv.org/abs/2510.10036}{{\ttfamily 2510.10036}}].

\bibitem{Culetu:2013fsa}
H.~Culetu, \emph{{On a regular modified Schwarzschild spacetime}},  \href{https://arxiv.org/abs/1305.5964}{{\ttfamily 1305.5964}}.

\bibitem{Culetu:2014lca}
H.~Culetu, \emph{{On a regular charged black hole with a nonlinear electric source}}, \href{https://doi.org/10.1007/s10773-015-2521-6}{\emph{Int. J. Theor. Phys.} {\bfseries 54} (2015) 2855} [\href{https://arxiv.org/abs/1408.3334}{{\ttfamily 1408.3334}}].

\bibitem{Simpson:2019mud}
A.~Simpson and M.~Visser, \emph{{Regular black holes with asymptotically Minkowski cores}}, \href{https://doi.org/10.3390/universe6010008}{\emph{Universe} {\bfseries 6} (2019) 8} [\href{https://arxiv.org/abs/1911.01020}{{\ttfamily 1911.01020}}].

\bibitem{Borges:2023fub}
H.A.~Borges, I.P.R.~Baranov, F.C.~Sobrinho and S.~Carneiro, \emph{{Remnant loop quantum black holes}}, \href{https://doi.org/10.1088/1361-6382/ad210c}{\emph{Class. Quant. Grav.} {\bfseries 41} (2024) 05LT01} [\href{https://arxiv.org/abs/2310.01560}{{\ttfamily 2310.01560}}].

\bibitem{Chandrasekhar:1985kt}
S.~Chandrasekhar, \emph{{The mathematical theory of black holes}}, Clarendon Press, Oxford (1985).

\bibitem{Kokkotas:1999bd}
K.D.~Kokkotas and B.G.~Schmidt, \emph{{Quasinormal modes of stars and black holes}}, \href{https://doi.org/10.12942/lrr-1999-2}{\emph{Living Rev. Rel.} {\bfseries 2} (1999) 2} [\href{https://arxiv.org/abs/gr-qc/9909058}{{\ttfamily gr-qc/9909058}}].

\bibitem{Berti:2009kk}
E.~Berti, V.~Cardoso and A.O.~Starinets, \emph{{Quasinormal modes of black holes and black branes}}, \href{https://doi.org/10.1088/0264-9381/26/16/163001}{\emph{Class. Quant. Grav.} {\bfseries 26} (2009) 163001} [\href{https://arxiv.org/abs/0905.2975}{{\ttfamily 0905.2975}}].

\bibitem{Porto:2016zng}
R.A.~Porto, \emph{{The Tune of Love and the Nature(ness) of Spacetime}}, \href{https://doi.org/10.1002/prop.201600064}{\emph{Fortsch. Phys.} {\bfseries 64} (2016) 723} [\href{https://arxiv.org/abs/1606.08895}{{\ttfamily 1606.08895}}].

\bibitem{Newman:1961qr}
E.~Newman and R.~Penrose, \emph{{An Approach to gravitational radiation by a method of spin coefficients}}, \href{https://doi.org/10.1063/1.1724257}{\emph{J. Math. Phys.} {\bfseries 3} (1962) 566}.

\bibitem{Kunduri:2013gce}
H.K.~Kunduri and J.~Lucietti, \emph{{Classification of near-horizon geometries of extremal black holes}}, \href{https://doi.org/10.12942/lrr-2013-8}{\emph{Living Rev. Rel.} {\bfseries 16} (2013) 8} [\href{https://arxiv.org/abs/1306.2517}{{\ttfamily 1306.2517}}].

\bibitem{Kehagias:2024yzn}
A.~Kehagias, D.~Perrone and A.~Riotto, \emph{{A short note on the Love number of extremal Reissner-Nordstr{\o}m and Kerr-Newman black holes}}, \href{https://doi.org/10.1016/j.physletb.2024.139109}{\emph{Phys. Lett. B} {\bfseries 859} (2024) 139109} [\href{https://arxiv.org/abs/2406.19262}{{\ttfamily 2406.19262}}].

\bibitem{bardeen1968}
J.M.~Bardeen, \emph{Non-singular general-relativistic gravitational collapse},  in \emph{Proceedings of the International Conference GR5}, (Tbilisi, USSR), p.~174, 1968.

\bibitem{Hayward:2005gi}
S.A.~Hayward, \emph{{Formation and evaporation of regular black holes}}, \href{https://doi.org/10.1103/PhysRevLett.96.031103}{\emph{Phys. Rev. Lett.} {\bfseries 96} (2006) 031103} [\href{https://arxiv.org/abs/gr-qc/0506126}{{\ttfamily gr-qc/0506126}}].

\bibitem{Feng:2026sra}
Z.-W.~Feng, H.-L.~Liu, Y.~Ling and Q.-Q.~Jiang, \emph{{Regular black hole with sub-Planckian curvature and suppressed exponential mass inflation}},  \href{https://arxiv.org/abs/2605.15576}{{\ttfamily 2605.15576}}.

\end{thebibliography}
\end{document}